\documentclass[journal]{IEEEtran}
\usepackage[english]{babel}
\usepackage{graphicx}

\ifCLASSINFOpdf
\else
\fi
\begin{document}

%

\title{Backward wave oscillator with resonator made of metal foils
(photonic BWO)}

\author{V.\,Baryshevsky, V.\,Evdokimov, A.\,Gurinovich,
E.\,Gurnevich, P.\,Molchanov}

\maketitle

\begin{abstract}
Numerical and experimental analysis of high power microwave
generation in photonic BWO, which uses foil photonic crystal is
presented. Single frequency excitation of the below cutoff modes
in the photonic BWO is analyzed and demonstrated.
\end{abstract}

\noindent \textbf{\small{photonic crystal, backward wave
oscillator (BWO), slow-wave structure (SWS)}}

\section{Introduction}

High power microwave generation by an electron beam in a waveguide
filled by a spatially periodic structure (metamaterial, photonic
crystal) is a subject for multiple studies both theoretical and
experimental
\cite{b6,b8,b9,b9+,b10,b11,b12,b13,b14,b15,b16,b17,b25}. A
photonic crystal could replace a corrugated waveguide or a
diffraction grating, which are usually used as a slow-wave
structure (SWS) in conventional {relativistic} backward wave
oscillators (BWOs), free electron lasers (FELs) and other types of
vacuum electronic devices.
The name ``photonic backward wave oscillator'' introduced in this
paper hereinafter means the BWO, which slow-wave structure is a
photonic crystal placed into a waveguide.

A conventional {relativistic} BWO consists of a spatially periodic
waveguide (slow-wave structure) into which a high-current electron
beam guided by a strong magnetic field is injected to interact
with the electromagnetic field inside the structure. The spatial
harmonics (modes) existing inside SWS could have oppositely
directed phase and group velocities, thus called backward waves.
The wave generated via electron beam interaction with  SWS travels
in the direction opposite to the beam velocity. The efficiency of
generation is determined by the beam-wave coupling coefficient,
which depends on the distance at which beam is guided inside the
slow-wave structure and the current density.

The photonic BWO enables some beneficial options, namely: use of
either pencil-like or sheet electron beams instead of annular one
and, thus, establishing the beam-wave interaction within the whole
waveguide cross-section. Additionally, the photonic crystals made
of foils, which are strained inside a cylindrical waveguide, were
shown \cite{b9} to support generation below cutoff of a
smooth-wall cylindrical waveguide. The same feature is expected
for photonic crystals made of threads, wires, rods or etc.

Operation in a waveguide below cutoff is an advantage of the
photonic BWO, because transverse dimensions of the generator can
be made much smaller than a wavelength.
This is important for miniaturization of low frequency microwave
generators (particularly, for L and S-band) \cite{b13}.

In present paper the photonic BWO, which uses foil photonic
crystal for high power microwave generation, is analyzed both
numerically and experimentally. Single frequency excitation of the
below cutoff modes in the photonic BWO is demonstrated.

\section{System description}
\label{sec1a}

\begin{figure}[htb]
\centering
\includegraphics[width=8cm]{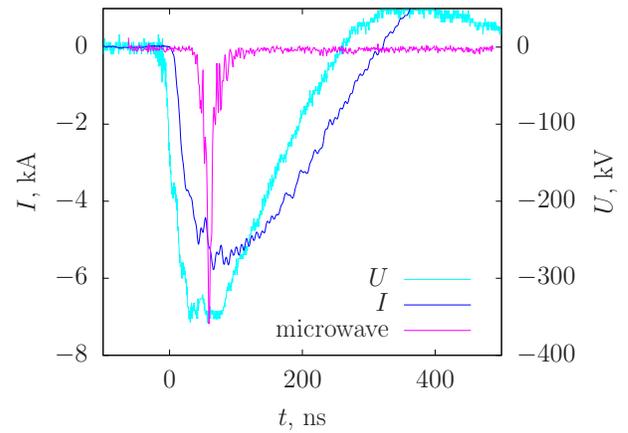}
\caption{Typical supplying voltage and beam current signals.}
\label{fig13}
\end{figure}

The photonic BWO is studied experimentally using a 350 keV, 5.5
kA, 150 ns electron beam, which is generated by a Marx generator:
{typical voltage and current pulses are presented in
Fig.\ref{fig13}}. The pencil-like electron beam of 40\,mm diameter
produced by 35\,mm graphite cathode is injected into SWS through
the mesh anode, which transparency is 77\%. The beam is guided by
an axial pulsed magnetic field, which is generated by the
discharge of a capacitor bank into a solenoidal coil of 600\,mm
length. The magnetic field pulse has amplitude about 1\,T and
duration of several milliseconds at half height. A synchronization
circuit ensures generation of the electron beam at the maximal
peak value of magnetic field.

\begin{figure}[htb]
\centering
\includegraphics[width=3.3 cm]{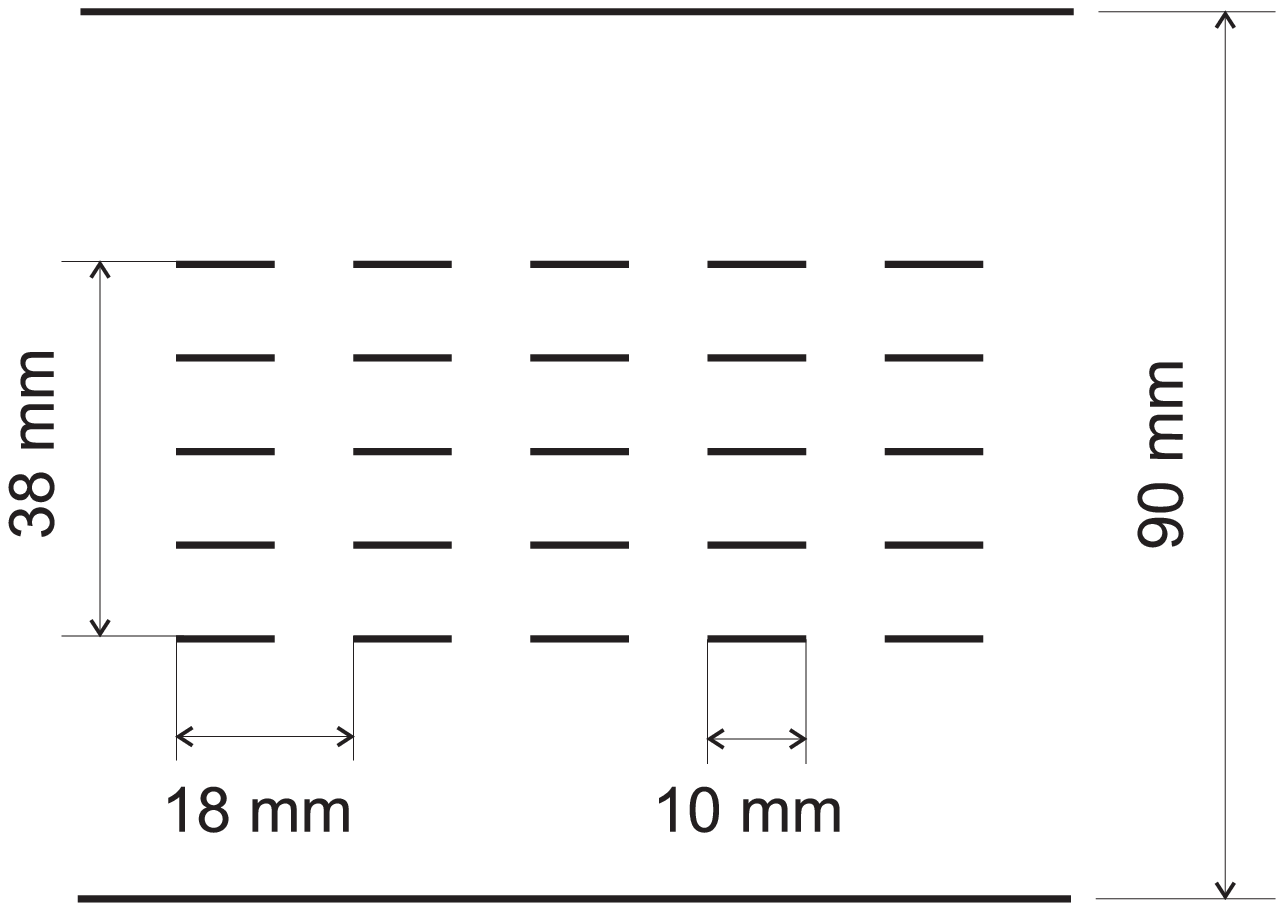}
\includegraphics[width=5 cm]{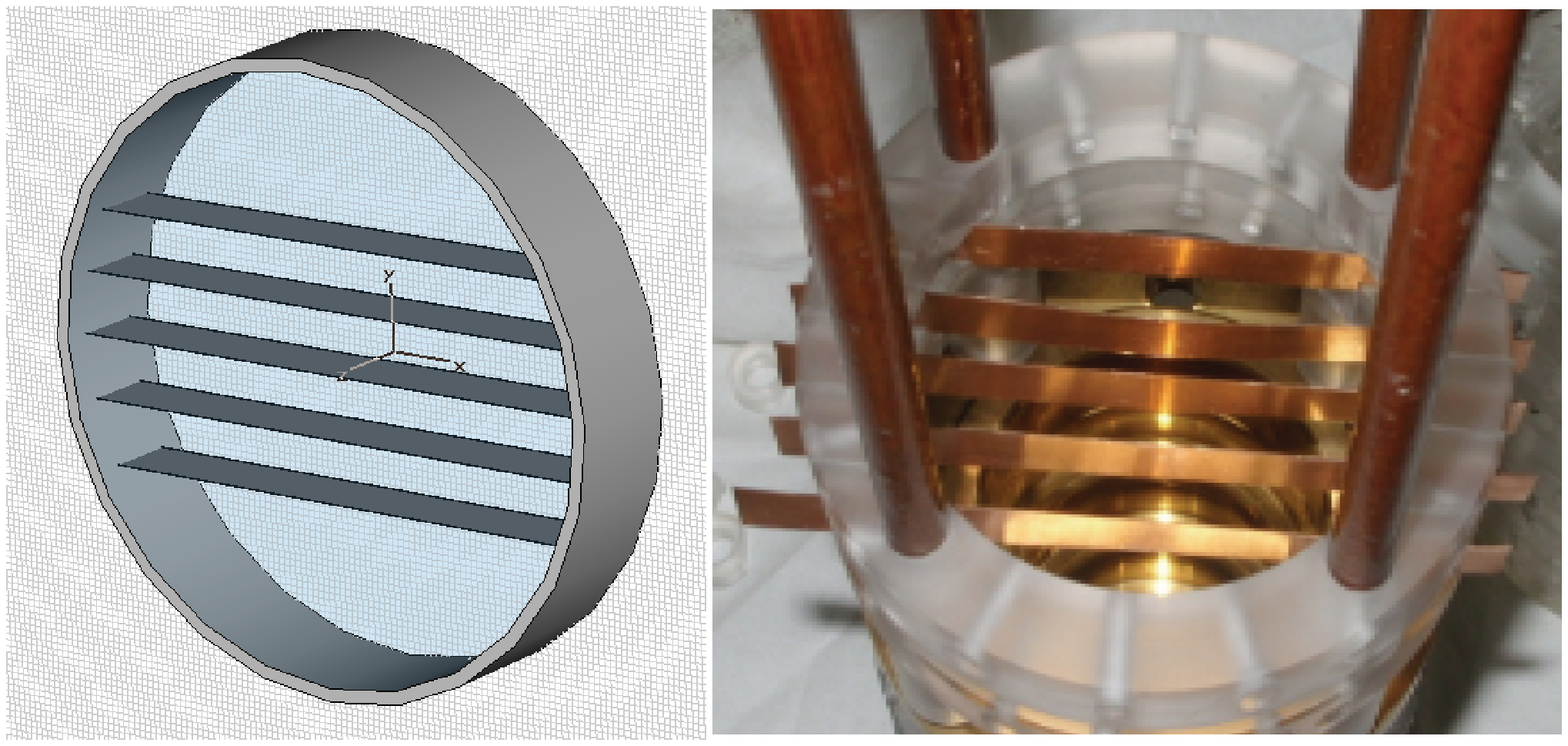}
\caption{Foil photonic crystal used in the experiments: the
side-view drawing (left), model (center) and foils installation
photo (right).} \label{fig1}
\end{figure}

The beam interacts with the foil photonic crystal, which is formed
by brass foils strained in the cross-section of cylindrical
waveguide with inner diameter 90\,mm. Foils' width and thickness
are 10~mm and 250~$\mu$m, respectively. The crystal comprises 10
periods, each period has 5 foils in cross-section (see
Fig.~\ref{fig1}).

A waveguide reflector is placed behind the anode to enable
reflection of the backward wave towards the output window.
The waveguide reflector consists of 4 periods of corrugated
waveguide. The inner diameter of the straight circular corrugated
waveguide is 50\,mm, corrugation period, width and depth are
29\,mm, 8\,mm and 5\,mm, respectively. Reflectance approaching
100\% is provided by this reflector in the range from 0 to
4.7\,GHz. The anode mesh enables effective reflection in the rest
frequency ranges.

Horn antenna with exit aperture 480~mm is used  for radiation
output. The strength of electric field is detected at the distance 5~m
by a receiving antenna and passed by the coaxial cable of RG\,402
type to the input of digital oscilloscope Tektronix TDS7404. The
detected time-domain waveforms are analyzed using short-time
Fourier transform to evaluate spectral content.

\section{Dispersion characteristics of foil photonic crystal}
\label{sec2}

Electromagnetic waves generation by an electron beam moving in a
photonic crystal can be analyzed using the dispersion equation,
which defines the relation between frequency and photon wave
number in the crystal.
The analytic solution of Maxwell equations for the photonic
crystal of the considered geometry cannot be found, thus the
dispersion properties of the crystal are analyzed numerically as
it was done in \cite{b9}.

\begin{figure}[ht]
\centering
\includegraphics[width=0.99\linewidth]{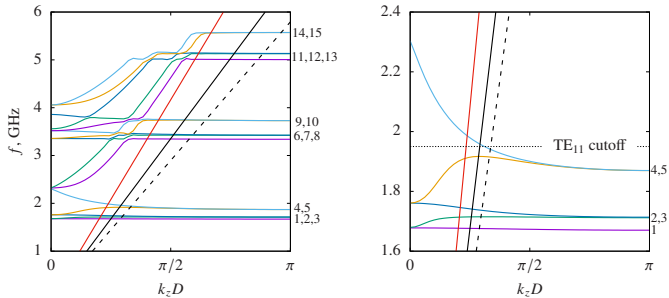}\\
\caption{Dispersion curves for low-order modes existing in the
foil photonic crystal; right plot shows more detailed view of
frequency range 1.6 - 2.4\,GHz. Beam lines for electron energy
200~keV (dashed black), 350~keV (solid black) and light line (red)
are shown.} \label{fig2}
\end{figure}

Dispersion curves for the low-order modes existing in the above
described photonic crystal are shown in Fig.~\ref{fig2}. The light
line $\omega=k_z c$ and the beam lines $\omega=k_z v_e$ for electron beam
energies 200 and 350\,keV are also shown.
Intersections of dispersion curves and beam lines correspond to
the points, where Cherenkov synchronism is achieved \cite{b23,
b24}.

Note that for some of modes shown in Fig.~\ref{fig2}, the waves
group velocity in these points is negative (i.e. is directed
backward the beam velocity). Such modes are classified in
\cite{b22} as negative dispersion ones.
Frequencies for these modes are below cutoff for the regular
cylindrical waveguide with the same inner diameter: for a
waveguide with internal diameter 90\,mm the cutoff frequency for
azimuthally symmetric mode TM$_{01}$ and for nonsymmetric mode
TE$_{11}$ are 2.55 and 1.95\,GHz, respectively.

\begin{figure}[htb]
\centering
\includegraphics[width=9cm]{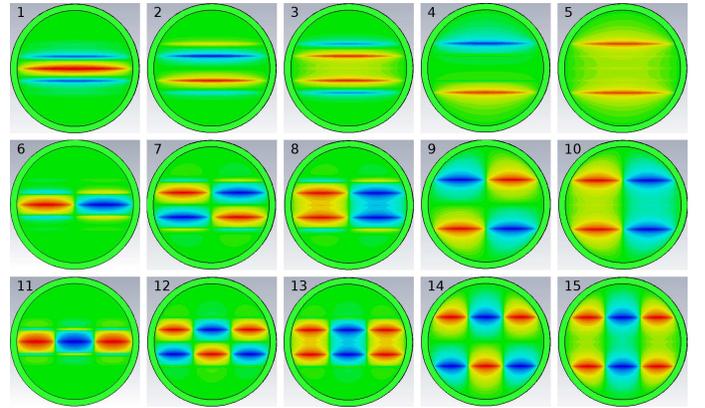}\\
\caption{E-field ($E_z$) patterns for the foil photonic crystal
dispersion modes shown in Fig. \ref{fig2} } \label{fig2b}
\end{figure}

The considered photonic crystal is not axially symmetric.
Therefore, one cannot establish a strict correspondence between
the types of electromagnetic waves, which are excited inside it,
and the eigenmodes of the regular cylindrical waveguide.
Structure of modes excited inside the foil photonic crystal for
$k_z$ values enabling Cherenkov synchronism is presented in
Fig.~\ref{fig2b}; mode numbering starts from the lowest mode, the
mode structure patterns are respectively numbered.
Field energy is mainly concentrated inside the crystal in between
the foils that makes these modes quite different from the
eigenmodes of the regular cylindrical waveguide.

The mode structure patterns (see Fig.~\ref{fig2b}) for modes
numbered 1,3,5,11 and 13 are symmetric with respect to vertical
and/or horizontal planes, while for modes 2,4,6,7,8,9,10,12,14 the
axial E-field component $E_z$ is distributed antisymmetric and
$E_z=0$ at the waveguide axis. Therefore, the electron beam
passing through the photonic crystal along the waveguide axis gets
into $E_z$ minima for antisymmetric modes and beam-wave
interaction for the them is suppressed as compared with symmetric
those. As a result, generation at 1.5 - 2.0\,GHz and 5\,GHz is
expected to be observed.


\section{Experimental results}
\begin{figure}[h]
\centering
\includegraphics[width=0.8\linewidth]{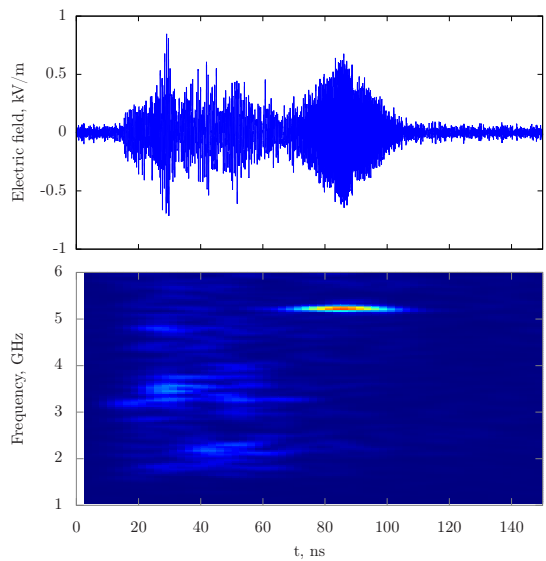}
\caption{Typical radiated time-domain waveform for 200 keV
electron beam \cite{b9} and its spectral content}
\label{fig2009stft}
\end{figure}

In experiments at electron beam energy 200\,keV, which are
described in paper \cite{b9},  generation at below cutoff
frequencies was observed, however, single frequency generation was
not obtained. In major part of shots the most of radiated power
density was concentrated in 5.2~GHz (see Fig.~\ref{fig2009stft})
i.e. in 11-th and 13-th modes, which structure is shown in
Fig.~\ref{fig2b}.

\begin{figure}[htb]
\centering
\includegraphics[width=0.8\linewidth]{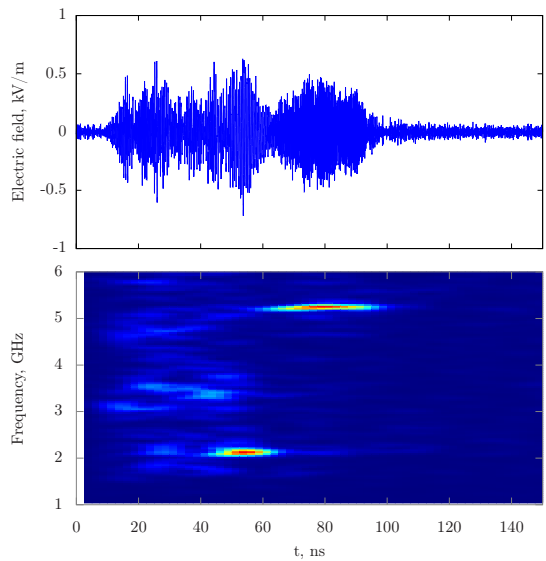}
\caption{Radiated time-domain waveform for 200 keV electron beam
and its spectral content for a shot, when below cutoff frequency
oscillations were excited.} \label{fig2009stft2}
\end{figure}

Cherenkov synchronism condition for these modes was fulfilled at
$k_z D$ value approaching $\pi$, enabling these modes to be
effectively excited by the 200-keV beam \cite{b9,b11}.
Symmetric mode structure enabled effective interaction with
electron beam over its whole cross-section. In some shots
generation at below cutoff frequencies was excited at rising diode
current (Fig.~\ref{fig2009stft2}). While generation for
antisymmetric modes is significantly suppressed that is in full
accordance with the above analysis.

The following reasoning justifies increase of electron beam energy
with the goal to observe single-frequency generation at below
cutoff modes.
According to Fig.~\ref{fig2} growth of the electron beam energy
shifts Cherenkov synchronism away from the $\pi$-point for modes 11 and 13,
thus excitation of these modes at electron beam energy 350~keV
is expected to be suppressed. Modes 1, 3 and 5 would contribute to
radiated power in 1.5--2.0~GHz.

\begin{figure}[h]
\centering
\includegraphics[width=0.8\linewidth]{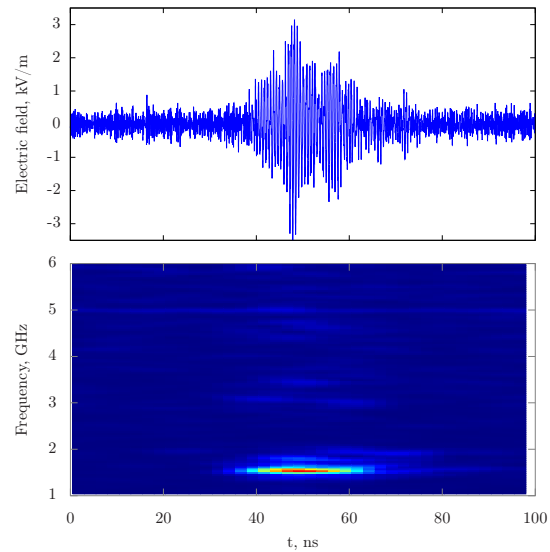}\\
\caption{Radiated time-domain waveform for 350 keV electron beam
and its spectral content} \label{fig2013stft}
\end{figure}

Example of experimentally observed at electron beam energy 350~keV
radiated waveform and their spectral content is shown in
Fig.~\ref{fig2013stft}. 
Generation at 11-th and 13-th modes is strongly suppressed (absent) in this experiment.
In Fig.~\ref{fig2013stft} the most of radiated power
density is observed at 1.55$\pm$0.05 GHz, which is below
cutoff.

Generation at this frequency can be associated with excitation of
the 1-st and/or the 3-rd modes (Fig.~\ref{fig2}). The detected
frequency 1.55\,GHz is a little bit lower than that (1.67\,GHz)
obtained from the analysis of dispersive curves for the photonic
crystal in the circular waveguide in the absence of electron beam.
This difference can be specified by influence of electron beam
space charge.



\section {Conclusion}
Single frequency generation at the below cutoff modes in the
photonic BWO, which uses foil photonic crystal for high power
microwave generation, is demonstrated. Excitation of either single
the lowest mode or several closely spaced lower modes results in
single frequency generation.


\end{document}